\documentclass[dvips]{article}
\usepackage{graphicx,amssymb,amsmath,times,wrapfig,units,cite,fancyhdr}
\usepackage{icrctc07}

\title{Observation of Seasonal Variations with the MINOS Far Detector}
\shorttitle{Observation of Seasonal Variations...}	

\authors{E.W. Grashorn$^{1}$,for the MINOS Collaboration$^{2}$}
\shortauthors{E.W. Grashorn et al.} 
\afiliations{$^1$ Univ. of Minnesota School of Physics \& Astronomy, 116 Church St.,
        Minneapolis, MN 55455, USA\\
        $^2$http://www-numi.fnal.gov/collab/collab.ps}

\abstract {An observation of seasonal variations in underground muon rate,
$R_{\mu}$, has
been performed at Soudan, MN, by the MINOS Far Detector.  The
four percent fluctuation seen over three years
was highly correlated to the temperature variations of the upper
atmosphere.  The
coefficient relating variations in temperature to variations in
muon rate was found to be: $\alpha_T = (T/R_{\mu})(\partial
R_{\mu}/ \partial T) = 0.87 \pm 0.03$, which is near the
expectation of 0.91.}
\begin{document}
\maketitle

\section{Introduction \& Motivation}\label{sec:intro}
When cosmic rays interact with molecules in the Troposphere, mesons
are produced which either interact again and
produce low energy cascades 
or decay into muons. 
While the temperature of the Stratosphere varies considerably within the day
in areas far from the equator, the temperature of the Troposphere
remains nearly constant, slowly changing over longer timescales such as
seasons.   Increases in
temperature of the Troposphere cause increases in volume and atmospheric
scale height, thus the height of the primary cosmic ray
interaction.  The higher in the atmosphere mesons are produced, the more time
they have to decay to muons, thus the rate of muons underground 
will increase as temperature increases 
\cite{Barrett:1952,Ambrosio:1997tc,Bouchta:1999kg}. 
Though this effect has been
measured by underground experiments, there has been little agreement
with the expectation.  

MINOS is a
long baseline neutrino oscillation experiment, with a $\nu_{\mu}$ beam
and Near Detector at Fermi National Accelerator Laboratory in Batavia,
IL.  The Far Detector is a 5.4~kt magnetized scintillator and steel tracking
calorimeter
located 720~m underground (2100~mwe) at the Soudan Underground Mine
State Park in Northern Minnesota.  Its depth, large acceptance and flat
overburden make it possible to observe cosmic-ray induced muons of
minimum surface energy \unit[0.7]{TeV} without
preference to direction, and thus detect the small seasonal
fluctuations in arrival rate.  The seasonal effect is enhanced as muon
energy increases, and the large size of the Far Detector allows a significant
accumulation of statistics with which to perform this analysis.
Additionally, the Far Detector has a magnetic field, which allows the
separation of particles by charge, so  MINOS will be the first experiment
to measure seasonal variations for $\mu^+$ separate from $\mu^-$. 
The consistency and availability of radiosonde 
temperature measurements from the NOAA IGRA (Integrated Global Radiosonde
Archive \cite{IGRA}) over the duration of the data set ensures a high
statistics 
temperature sample as well, increasing the probability of a positive
correlation.  The data used in this analysis were collected
over three years, from 1~August~2003 - 1~August~2006 for three complete
cycles, numbering 24~million muons.
The relationship
between the temperature and intensity can be
expressed as~\cite{Barrett:1952}:
\begin{equation}\label{eq:alpha} 
\frac{\Delta I_{\mu}}{I^{0}_{\mu}} =
\int_{0}^{\infty}dX\alpha(X)\frac{\Delta T(X)}{T(X)}
\end{equation}
where $\Delta I_{\mu}$ are the fluctuations about $I^{0}_{\mu}$.
The short lived mesons produced in the upper atmosphere interact or decay as they descend toward the earth. 
The meson decay channels result in muons with nearly the same
energy as the parent meson, while interactions produce lower energy
cascades that are filtered by the rock overburden above the Far Detector.
These outcomes are energy dependent, separated by the
``critical energy''~\cite{Ambrosio:1997tc}.
The ``Effective Temperature'' ($T_{eff}$)
approximates the upper atmosphere as an isothermal body,
weighting the temperature of the pressure levels to have a
uniform amount of matter.  In the $\pi$ scaling limit, $T_{eff}$ is~\cite{Ambrosio:1997tc}: 
\begin{equation}\label{eq:teff}
T_{eff}=
\frac{\int\frac{dX}{X}T(X)\Big(e^{\frac{-X}{\Lambda_{\pi}}}-e^{\frac{-X}{\Lambda_{N}}}\Big)}{\int
\frac{dX}{X}\Big(e^{\frac{-X}{\Lambda_{\pi}}}-e^{\frac{-X}{\Lambda_{N}}}\Big)},
\end{equation}
where X is the scale height of the atmosphere,  $\Lambda_{N} =
\unit[120]{gm/cm^{2}}$ and
$\Lambda_{\pi} =  \unit[160]{gm/cm^{2}}$ are the nucleon and pion atmospheric attenuation
lengths, respectively.  For a detector counting discrete particles, the intensity is
written $I_{\mu} = R_{i}/\epsilon A_{eff}\Omega$, where $R_i = N_i/t_i$,
the number of muons observed over time $t_i$,$A_{eff}$ is the effective
area, $\epsilon$ is the efficiency, and $\Omega$ is the solid angle
observed.  Every term but the rate is constant over time, so:  $\frac{\Delta I_{\mu}}{I^{0}_{\mu}} = \frac{\Delta
R_{\mu}}{\left<R_{\mu}\right>}$.  With these definitions and eq.
\ref{eq:alpha}, we can write the experimental
determination of $\alpha_T$:  
\begin{equation}
\int_{0}^{\infty}dX\alpha(X)\frac{\Delta T(X)}{T(X)}=
\alpha_T\frac{\Delta T_{eff}}{<T_{eff}>} =  \frac{\Delta
R_{\mu}}{<R_{\mu}>}. 
\end{equation}
\section{The Data}
The data for this analysis were accumulated over a three year span,
beginning on 1 August, 2003,
%
at a time when the detector was fully
operational.
Beginning with 40.3 million cosmic ray tracks, a series of
cuts were performed \cite{Mufson:2007}.  
Pre-analysis cuts include: failure of
demultiplexing figure of merit, multiple muon (multiple muons aren't
included in the Monte Carlo), ``bad run'' and bad
magnet coil status. 
Analysis cuts include: track length less than 2~m
number of planes less than 20, $\chi^2_{reco}>1.0$ and either track vertex
or end point outside of the fiducial volume of the detector. 
A total of
24.7 million events survived these cuts for the combined sample.   
$T_{eff}$ was found using weather data from
International Falls, MN weather station.  Balloon flights were usually done
twice a day, with the maximum height reached at noon and midnight, and sampled temperatures from at least six different
heights.  Days in which there were not exactly two temperature readings
or that both measurements did not reach a column depth of at least
$\unit[60]{gm/cm^2}$ were excluded from the data set.

\section{Analysis}
\begin{figure}
\begin{center}
\includegraphics[width=.49\textwidth]{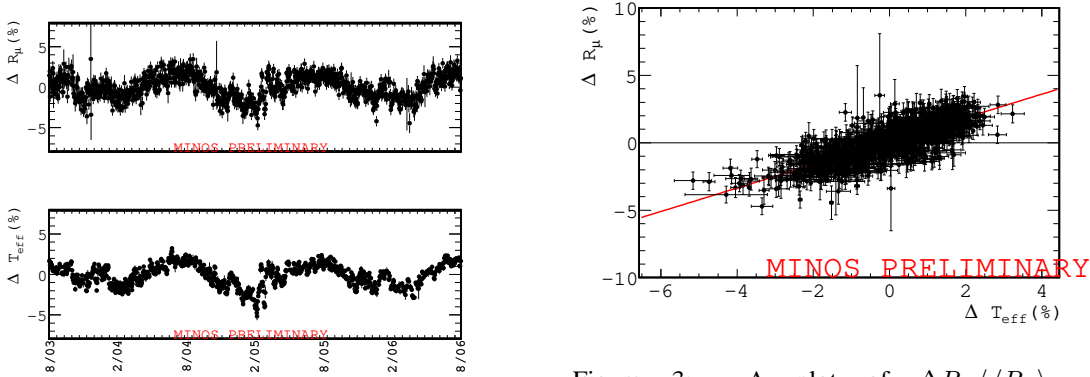}
\vspace{-20pt}
\caption{\label{fig:seasonal}$\Delta R_{\mu}$, (top), and $\Delta T_{eff}$ (bottom) from 8/03-8/06, binned by day.}
\vspace{-15pt}
\end{center}
\end{figure}

\begin{figure}[h]
\begin{center}
\vspace{-10pt}
\includegraphics[width=.49\textwidth]{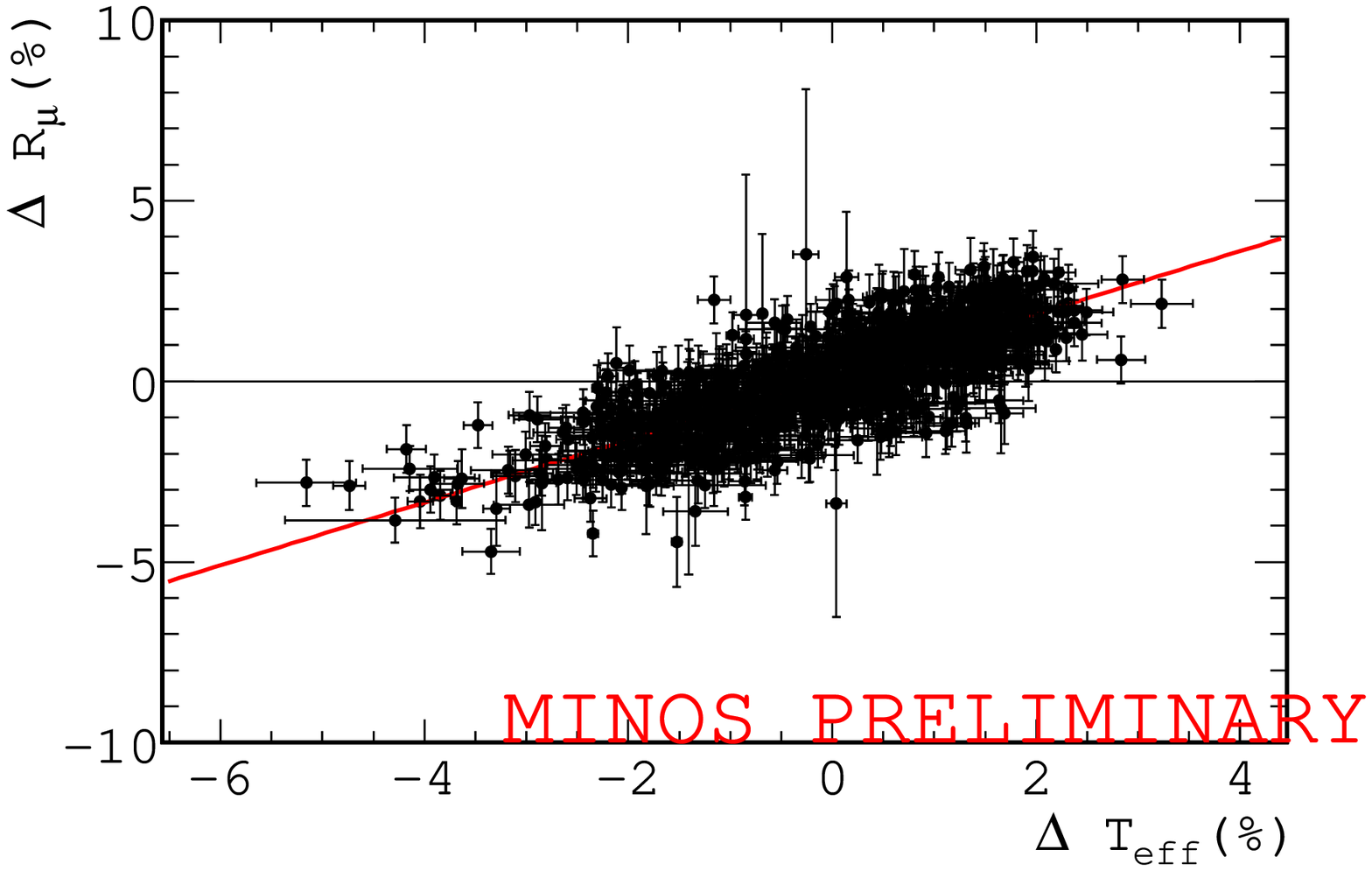}
\vspace{-20pt}
\caption{\label{fig:RTCorrP} A plot of $\Delta R_{\mu}/\left<R_{\mu}\right>$ vs. $\Delta T_{eff}/\left<T_{eff}\right>$ for
single muons.  The fit $\chi^2/ndof =
1420/953$, and correlation coefficient R = 0.79.}
\end{center}
\vspace{-20pt}
\end{figure}

  Upon examination of the data, it was found that on four days there were
fluctuations that deviated in an erratic manner.   The great stability
of the detector over the 1096 days of data and the fact that they were documented by 
the Control Room Logbook made these days stand out
and diagnose as hardware issues. 
To find the rate for each day, the number of muons counted was
divided by that day's livetime.
 $T_{eff}$ was
calculated for two times each day using the IGRA temperature data and
\ref{eq:teff}, and the error was found by $\sigma^2 =
\left<T_{eff}^2\right>-\left<T_{eff}\right>^2$ 
added in quadrature with $0.05^{\circ}$.
\begin{figure}
\begin{center}
\includegraphics[width=0.48\textwidth]{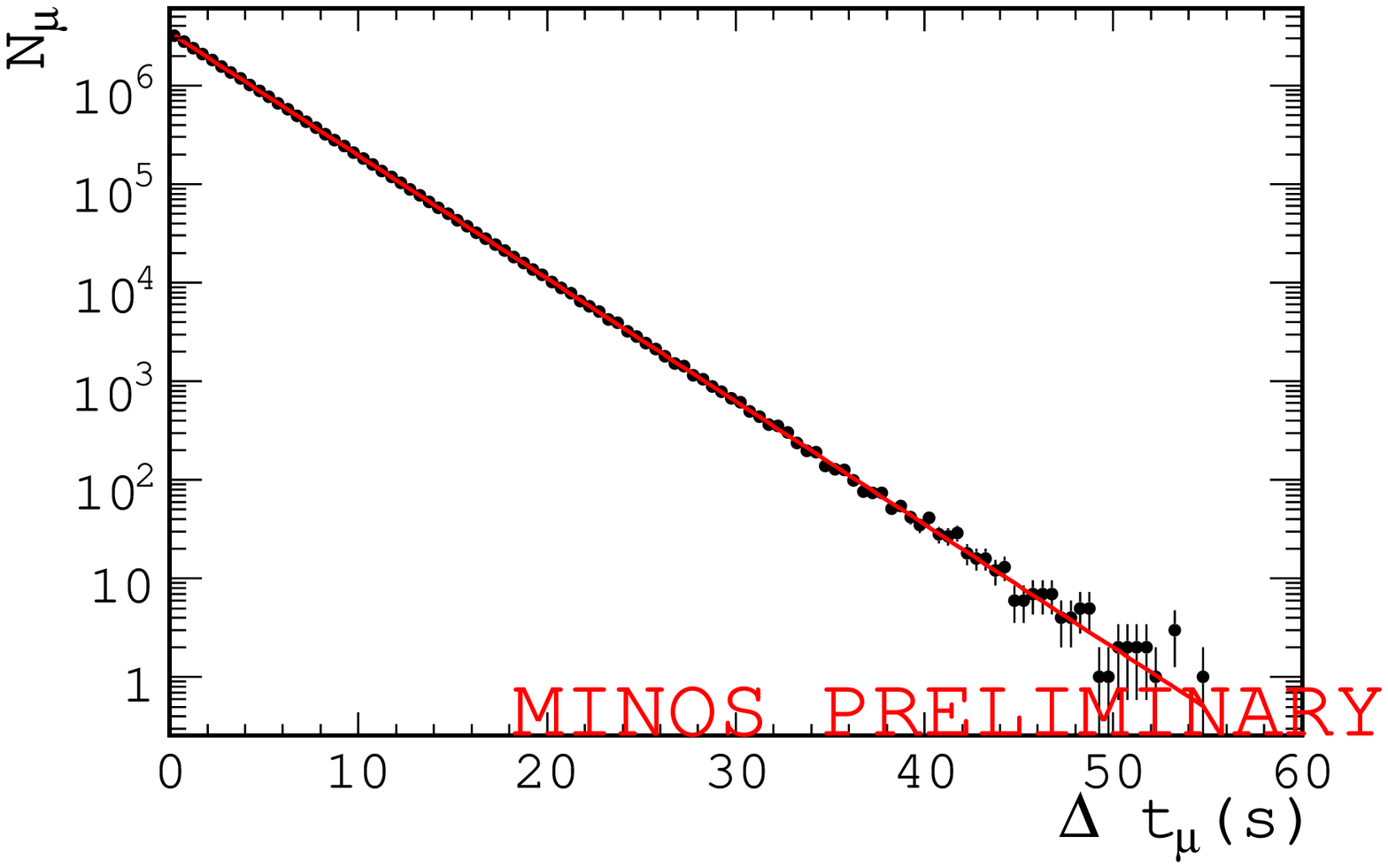}
\end{center}
\vspace{-20pt}
\caption{\label{fig:dt}Time between consecutive $\mu$
arrivals in log y.  A Poisson fit gives
$\chi^2/ndof = 90.87/105$; $\left<R_{\mu}\right>$ (from slope) = 
 $0.2872\pm0.0001$}
\vspace{-20pt}
\end{figure}
The fit results from Fig.~\ref{fig:dt} was
used to find $\left<R_{\mu}\right>$ over three years, 0.287~Hz.  
Histograms of the
deviations from the mean for both  $R_{\mu}$ and  $T_{eff}$ are shown in
Fig.~\ref{fig:seasonal}, binned by day.  
The expected periodic fluctuation in $T_{eff}$,
with maxima in July, minima in January, is very clearly shown, as is a
very similar (nearly indistinguishable) fluctuation in $R_{\mu}$. An
independent analysis used a smoothed time series, and their results were highly consistent with what is shown here.  To quantify the correlation between rate and temperature, a plot of
$R_{\mu}(T_{eff})$ was produced (Fig.~\ref{fig:RTCorrP}) and a linear
regression was fit using ROOT's MINUIT fitting package.  This package
accounts for error bars on both the x and
y axis using a numerical minimization method. This fit gives
 $\alpha_T = 0.87 \pm 0.03$ from the slope.
%
\begin{figure*}
  \begin{center}
\vspace{-10pt}
 \includegraphics[width = .8\textwidth]{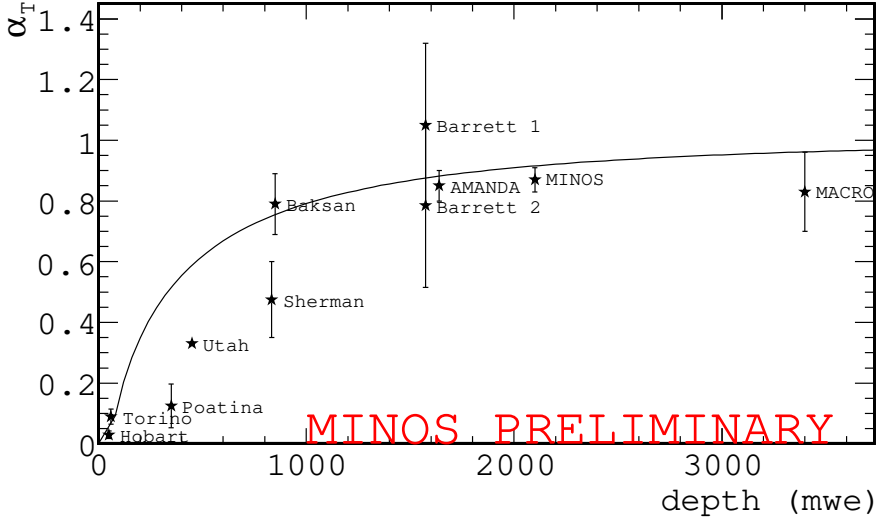}
\vspace{-20pt}
\caption{\label{fig:alphaGlobal}  The theoretical $\alpha_{T}(X)$ (solid
curve) for slant depths up to 4000 mwe.  
The MINOS point is from this analysis, 
Barrett 1,2~\cite{Barrett:1952}, AMANDA~\cite{Bouchta:1999kg}; all other points from~\cite{Ambrosio:1997tc}}
 \vspace{-20pt}	
\end{center}
\end{figure*}
In order to compare our experimental $\alpha_T$ to the theoretical
expectation, a simple numerical program was written to find the expected
value given by~\cite{Barrett:1952}:
\begin{equation}
\left<\alpha_{T}\right>_{\pi}=\Big<1+\frac{\gamma}{(\gamma+1)}\times\frac{\epsilon_{\pi}}{1.1E_{th}cos\theta}\Big>
\end{equation}
Note that this expression is only valid for pions.  Future work will
involve this kaon contribution, which should lower the expected
$\alpha_T$ since kaons are short lived and always decay.
A muon energy and $\cos\theta$ were chosen out of the differential
muon intensity~\cite{Gaisser:1990vg}, 
\begin{equation}\label{eq:dI}
\frac{dI_{\mu}}{dE_{\mu}} =
0.14E_{\mu}^{-(\gamma+1)}\Big[\frac{1}{1+1.1 E_{\mu}\cos\theta/ \epsilon_{\pi}}\Big]
\end{equation}
where $\gamma=1.7$ is the muons spectral index \cite{Mufson:2007}.
a random
azimuthal angle, $\phi$ was chosen and combined with $\cos \theta$ and
 Soudan rock overburden map \cite{Goodman:2007} to
find the slant depth.  The threshold surface energy
required for a muon to survive this column depth is found from
$E_{th}(\theta,\phi) = a(e^{bX}-1)$, where a = 0.45 TeV and $b =
\unit[0.44]{[kmwe]^{-1}}$ for Soudan rock \cite{Mufson:2007}, column depth
$X=X(\theta,\phi)$,  and
if the chosen $E_{\mu}>E_{th}$, it was used in the calculation of the
theoretical $\left<\alpha_T\right>_{\pi}$.  This was repeated for 10,000 successful
$E_{\mu}$ to find $\left<\alpha_T\right>_{\pi} = 0.91$ for MINOS, which is very
near to the experimental value, $0.87\pm0.03$.  To compare the MINOS
result with other underground experiments, this process was
repeated for standard rock ($a = \unit[0.50]{TeV}$ and $b = \unit[0.4]{[kmwe]^{-1}}$), flat
overburden, and $X = H /\cos\theta$, where H is the detector depth in
mwe, using 10,000 successful muons at depths from 0 to 4,000 mwe.  The
result of this calculation, along with data from other experiments, can
be seen in Fig.~\ref{fig:alphaGlobal}. The MINOS result matches the
expectation and has tighter error bars than both recent results, AMANDA~($\pm0.05$~\cite{Bouchta:1999kg}) and MACRO~($\pm0.13$~\cite{Ambrosio:1997tc}).

The curvature of the
track is used to determine the momentum and charge of the particle, so a
charge sign confidence cut was required.
This cut was charge over momentum divided by the error in the
determination of charge over momentum ($\frac{q/p}{\sigma_{q/p}}>2.2$),
determined from previous investigations of the muon charge ratio. 
\begin{figure}[h]
\begin{center}
\includegraphics[width=.48\textwidth]{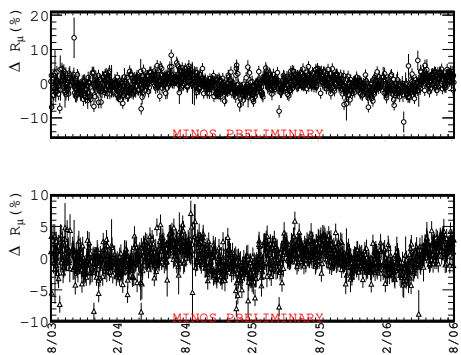}
\vspace{-20pt}
\caption{\label{fig:PercentDiffN}$\Delta R_{\mu}/\left<R_{\mu}\right>$
for $\mu^+$ (open triangles, top) and $\mu^-$ (open
circles, bottom), binned by day. }
\vspace{-10pt}
\end{center}
\end{figure} 
That left 8.8 million events; 5.1 million positive,
3.7 million negative, 
which is consistent with the published MINOS charge
ratio.  Fig.~\ref{fig:PercentDiffN}(t) shows $\Delta R_{\mu^+}$ (open
triangles) and Fig.~\ref{fig:PercentDiffN}(b) shows $\Delta
R_{\mu^+}$(open circles) over the same time period, binned by day.  The
sample of muons is 
smaller than for the $\mu_{tot}$ sample, thus the error bars on
$R_{\mu}$ are larger, but the trade off is that the error bars on the
temperature are much smaller since the small fluctuations over several
days are not washed out.  Performing the same fit of $R_{\mu}(T_{eff})$
as for the $\mu_{tot}$ sample on $\mu^+$ separate from $\mu^-$
resulted in a slope of $0.845 \pm 0.036$  and $0.843 \pm 0.042$
respectively.  These correspond highly to each other, and are within one
sigma of $\alpha_T$ found from the $\mu_{tot}$ sample.

\section{Conclusions}
A three year sample of 42 million cosmic ray induced muons has been
collected by the MINOS Far Detector and daily rate fluctuations have
been compared to daily fluctuations in atmospheric temperature, and
these distributions were shown to be highly correlated, with a
correlation coefficient of 0.79.  The constant of proportionality
relating the two distributions, $\alpha_T$, was found to be
$0.87\pm0.03$, which, within the error band, is in good agreement with the theoretical
expectation in the pion-only approximation
of $\left<\alpha_T\right>_{\pi} = 0.91$.  This suggests that the majority of
muons seen in the Far Detector were generated by pion parents.  
\section{Acknowledgments}
This work was supported by the U.S. Department of Energy and the
University of Minnesota. 

\bibliography{SeasonalBP}
\end{document}